\documentclass[journal,10pt]{IEEEtran}

\usepackage{graphicx}
\usepackage[cmex10]{amsmath}
\usepackage[utf8]{inputenc}
\usepackage{graphicx}
\usepackage{array}
\usepackage{floatrow}
\usepackage{caption}
\usepackage{subcaption}
\usepackage{gensymb}
\usepackage{url}
\usepackage{balance}
\usepackage{enumerate} 
\usepackage{amsbsy}
\usepackage[normalem]{ulem}
\usepackage{textcomp}
\usepackage{siunitx}
\usepackage{balance}
\usepackage{placeins}
\usepackage{tabularx, ragged2e}
\usepackage{listings}
\usepackage[dvipsnames]{xcolor}
\usepackage{svg}
\usepackage{amssymb}
\usepackage{stfloats}
\usepackage{subcaption}
\usepackage{booktabs}
\usepackage{multirow}
\usepackage{adjustbox}
\usepackage{tabularx}
\usepackage{pifont}
\usepackage{makecell}

\usepackage{hyperref}[hidelinks]

\usepackage[compact]{titlesec}
\titlespacing*{\subsection}{0pt}{0.8ex plus .2ex}{0.3ex}

\usepackage{cleveref}

\def\BibTeX{{\rm B\kern-.05em{\sc i\kern-.025em b}\kern-.08em
    T\kern-.1667em\lower.7ex\hbox{E}\kern-.125emX}}
    
\usepackage{xspace}

\newcommand{\iPerf}{$\tt iPerf$\xspace}
\newcommand{\FFmpeg}{$\tt FFmpeg$\xspace}

\DeclareTextFontCommand{\textcomputer}{\fontfamily{cmr}\selectfont}

\begin{document}

\title{Experimental Insights into UDP-Based Video and Control Traffic over IEEE 802.11p ITS-G5}

\author{Antonio Solida, Gaetano Orazio Cauchi, Salvatore Iandolo, Marco Savarese,
        \\Martin~Klapez,
        Maurizio~Casoni,
        and Carlo~Augusto~Grazia\\% <-this % stops a space
\textit{\{name.surname\}@unimore.it}
\thanks{A. Solida, G.O. Cauchi, S. Iandolo, M. Savarese, M. Klapez, M. Casoni, and C. A. Grazia are with the Department of Engineering \emph{Enzo Ferrari}, University of Modena and Reggio Emilia, via Pietro Vivarelli, 10, 41125, Modena, Italy.
\\
This study was carried out within the MOST – Sustainable Mobility National Research Center and received funding from the European Union Next-GenerationEU (PIANO NAZIONALE DI RIPRESA E RESILIENZA (PNRR) – MISSIONE 4 COMPONENTE 2, INVESTIMENTO 1.4 – D.D. 1033 17/06/2022, CN00000023). This manuscript reflects only the authors’ views and opinions, neither the European Union nor the European Commission can be considered responsible for them.
\\}% <-this % stops a space
}

\IEEEoverridecommandlockouts
\IEEEpubid{\makebox[\columnwidth]{979-8-3315-6420-9/26/\$31.00~\copyright~2026 IEEE \hfill}
\hspace{\columnsep}\makebox[\columnwidth]{ }}

\maketitle
\IEEEpubidadjcol

\pagestyle{plain}
\pagenumbering{gobble}

\begin{abstract}

Vehicular applications such as cooperative driving, teleoperation, and real-time perception increasingly rely on low-latency wireless communication. In this context, ITS-G5, based on IEEE 802.11p, represents a key technology for enabling direct vehicle-to-vehicle and vehicle-to-infrastructure communication. Despite its relevance, experimental studies focusing on the performance of UDP-based traffic over IEEE 802.11p under realistic conditions remain limited.

This paper presents an experimental evaluation of UDP transmission over an IEEE 802.11p ITS-G5 testbed composed of Raspberry Pi–based onboard units and commercial roadside units. The analysis investigates the impact of different modulation and coding schemes (MCS). It also evaluates two network-layer configurations (IPv4 unicast and IPv6 multicast) and the use of CAKE for active queue management. In addition to synthetic traffic generated with iPerf, the evaluation includes real-time video streaming using MPEG-TS over UDP to emulate latency-sensitive vehicular applications. Results show that the modulation scheme is the dominant factor influencing latency at low traffic loads, while the choice of transmission mode and IP version becomes increasingly significant under congested conditions. Higher-order modulations significantly reduce latency and variability, whereas IPv6 multicast exhibits greater delay dispersion than IPv4 unicast. Furthermore, active queue management does not seem to improve delay predictability. These findings provide practical insights for configuring ITS-G5 networks supporting latency-sensitive vehicular services.
\end{abstract}

\begin{IEEEkeywords}
IEEE 802.11p, Raspberry Pi, UDP, Video streaming, MCS, teleoperation, remote driving.
\end{IEEEkeywords}

\section{Introduction}
\label{sec_intro}

The evolution of vehicular technologies and the ever-increasing demand for reliable communication services have spurred significant research and development in recent years. Low-latency wireless connectivity is a key requirement for modern vehicular applications. Among these, ITS-G5 stands out as a key technology for enabling direct vehicle-to-vehicle (V2V) communications, which is particularly crucial for applications such as cooperative platooning and remote operations. While ITS-G5 has been a prominent technology for V2V communication, the ongoing exploration of LTE-V2V solutions and comparisons with IEEE 802.11-based technologies continue to drive research in this area. However, experimental evaluation of LTE-V2V remains limited by the scarcity of readily available commercial hardware. Even with the maturity of ITS-G5, identifying optimal protocols and mechanisms to effectively support diverse vehicular applications remains an active area of investigation.

\begin{figure}[t]
    \centering
    \includegraphics[width=0.75\textwidth]{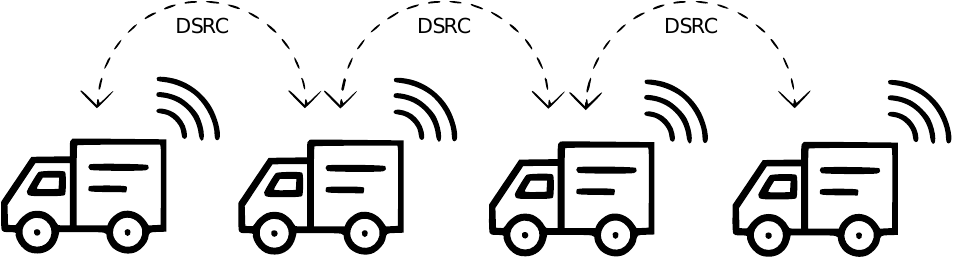}
    \caption{Vehicles Platooning.}
    \label{fig:platooning}
\end{figure}

\begin{figure}[t]
    \centering
    \includegraphics[width=1\textwidth]{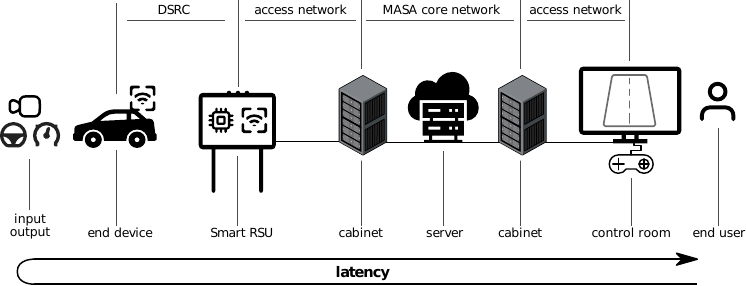}
    \caption{Remote driving \& Teleoperation.}
    \label{fig:remote}
\end{figure}

Many critical vehicular applications rely on UDP-based single-hop communications, particularly those that demand real-time data streams with stringent latency requirements. This is particularly relevant to emerging use cases, such as cooperative platooning of heavy-duty vehicles, which enable coordinated maneuvers, enhance safety, and reduce fuel consumption \cite{7286902} \cite{yang2022autonomous} (Figure \ref{fig:platooning}). Furthermore, remote driving and teleoperation \cite{solida2025remote} applications benefit from UDP's low overhead and reduced transmission delay, potentially enabling remote control of vehicles in remote or hazardous environments (Figure \ref{fig:remote}). These applications are highly sensitive to network characteristics and require careful characterization of the communication channel to ensure reliable performance under realistic operating conditions.

A key factor enabling continued research is the availability of support for ITS-G5 within the Linux ecosystem, which enables researchers to experimentally evaluate this technology using both open-source and custom-built Roadside Units (RSUs) and limited commercial RSUs. A significant advantage of ITS-G5 is its ability to enable immediate data exchange without the complex association and authentication procedures often required in infrastructure-based Wi-Fi networks, thereby reducing communication latency.

This paper presents a performance evaluation of ITS-G5 in a real testbed, analyzing generic UDP traffic and UDP streams carrying MPEG-TS video; it leverages both open-source hardware platforms (specifically utilizing our self-made OBU \cite{iandolo2025odu}) and a combination of commercial and custom-designed RSUs. This approach allows evaluation of ITS-G5's ability to support V2V applications.

The paper is organized as follows.
Section~\ref{sec_related} provides an overview of related works.
Sections~\ref{sec_hardware} and \ref{sec_testbed} present the hardware platform and its configuration.
Section~\ref{sec_test_results} presents the tests performed and analyzes the results, while
Section~\ref{sec_conclusions} concludes the article.

\section{Related Works}
\label{sec_related}

Existing literature on IEEE 802.11p-based vehicular communications has primarily focused on physical-layer aspects, such as channel behavior, modulation schemes, and link reliability, or on connection-oriented transport protocols such as TCP. In contrast, experimental work analyzing UDP traffic remains limited, despite its critical relevance to latency-sensitive applications.

Regarding transport-layer performance in vehicular networks, Grazia et al. \cite{grazia2018performance} presented the first experimental analysis of TCP over OCB (Open Control Buffer) mode in a Linux environment. Their study, conducted on a real testbed using the 5.9 GHz band with 10 MHz channels, identified suitable TCP configurations for reducing latency in vehicular scenarios. Similarly, Tahir et al. \cite{tahir2020vehicular} evaluated transport-layer performance in road weather service scenarios by comparing ITS-G5 with 5G infrastructure. Their experiments on a real test track demonstrated that, while ITS-G5 is optimized for vehicular environments, 5G networks currently offer superior reliability and throughput. These works highlight the challenges of managing latency and packet loss in vehicular networks but do not extensively address UDP-specific behaviors under demanding conditions.

In the domain of latency-critical video streaming, Pereira et al. \cite{pereira2018vehicle} proposed a V2V video transmission system based on IEEE 802.11p/WAVE to support driver decision-making during overtaking maneuvers. Their system, utilizing onboard cameras and encoders to transmit real-time video streams between vehicles, was evaluated in urban and highway scenarios. The results indicated that while end-to-end delay increases with vehicle speed due to larger inter-vehicle distances, the system remains feasible for assisted-driving applications. Earlier work by Rene et al. \cite{rene2011analysis} investigated video streaming quality on highways, finding that although fast handoff techniques mitigate blackouts caused by mobility-induced network reconfiguration, they impose significant limitations on service viability.

At the MAC layer, Høiland-Jørgensen et al. \cite{hoiland2017ending} addressed the fundamental tension between airtime fairness and low latency in IEEE 802.11 networks. They demonstrated that conventional queue management strategies lead to buffer bloat and throughput anomalies in heterogeneous traffic scenarios. Their work introduced the CAKE scheduler, which combines fair queuing with active queue management (AQM) to achieve per-flow latency control and equitable airtime allocation. Experimental results showed significant reductions in queuing delay, establishing CAKE as an effective discipline for variable-rate links. These findings are directly relevant to ITS-G5 deployments, where the coexistence of safety-critical CAM messages and high-bandwidth video streams challenges traditional MAC-layer designs.

Finally, regarding the configuration of Modulation and Coding Schemes (MCS), Daldoul et al. \cite{daldoul2018ieee} analyzed the impact of regulatory power constraints on high-throughput WLAN technologies (e.g., 802.11n/ac). Their study revealed that channel bonding and spatial multiplexing, while increasing peak data rates, can reduce communication range and diversity gain. They proposed a rate-ordering scheme to improve throughput stability in such conditions. For selecting MCS levels under varying channel conditions, this work relies on the systematic analysis presented in \cite{7906621}. That study characterizes the trade-off between robustness and spectral efficiency, offering insights into how different MCS settings impact throughput, range, and reliability. We align our MCS configurations with the methodology reported in that reference to ensure consistency with established performance analyses.

\section{Hardware Setup}
\label{sec_hardware}

\begin{figure}[h!]
    \centering
    \includegraphics[width=0.6\textwidth]{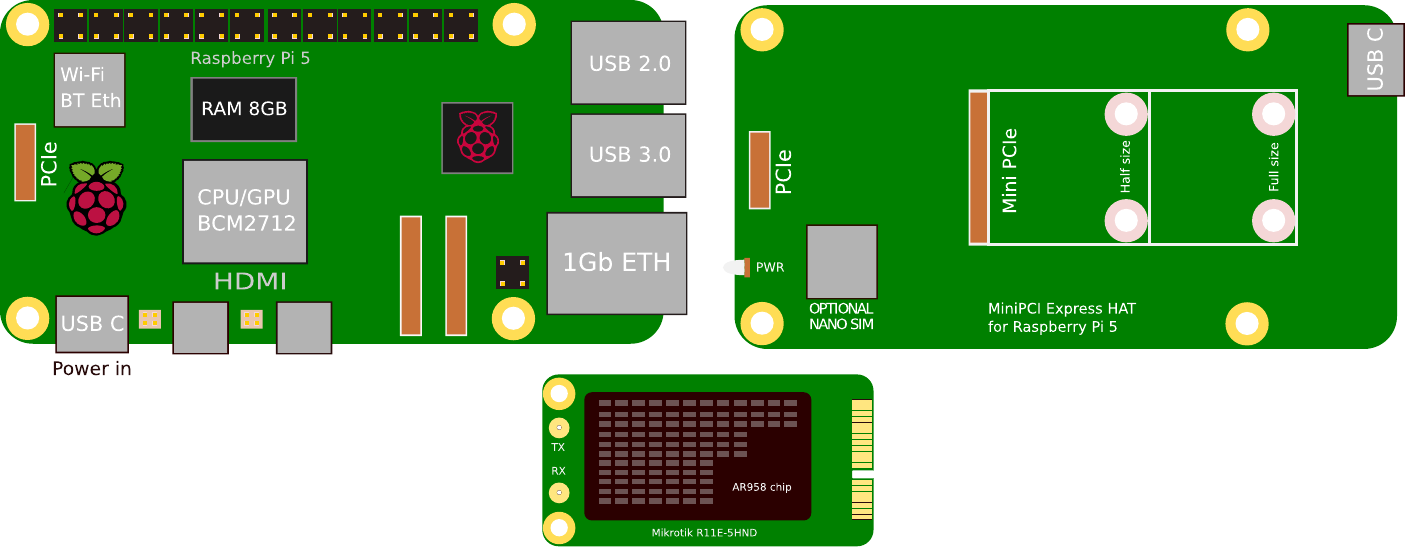}
    \caption{MiniPCI Express HAT | AR9580 Wi-fi card.}
    \label{fig:rasp_hat}
\end{figure}

The evaluation process used readily available development platforms, with a focus on balancing processing capabilities, expandability, and overall cost. The Raspberry Pi 5 was selected as the primary platform for its PCIe 2.0 interface, which enables the integration of external wireless adapters. This board incorporates an Arm Cortex-A76 processor, 8 GB of LPDDR4 memory, and 128 GB of flash storage, powered by a USB-C supply rated at 5 V and 5 A. As detailed in Figure \ref{fig:rasp_hat}, a Mikrotik R11E-5HND wireless card, utilizing a Qualcomm Atheros AR9580 chipset, was interfaced via the PCIe slot; this was necessary because the integrated network interface does not support the 5.9 GHz ITS band. The AR9580 operates within the 5.9 GHz ITS band and is supported by a modified Linux ath9k driver, enabling IEEE 802.11p V2X functionality; it has also enabled debug mode, the only state-of-the-art mode that allows certain transmission parameters, such as MCS, to be modified\footnote{https://github.com/AlfaSierra92/DCC-over-802.11p}. Communication is established in OCB (Outside the Context of a BSS) mode, as defined by the IEEE 802.11p standard, enabling direct peer-to-peer communication without requiring access point association—a mode specifically designed for vehicular and ad hoc environments, offering a nominal peak throughput of 27 Mbit/s.

The system's video acquisition is performed using a Raspberry Pi Camera Module 3, a compact imaging device that incorporates a Sony IMX708 12-megapixel sensor; this camera delivers strong performance in low-light conditions, supports HDR, and incorporates autofocus to ensure sharp images across a range of distances. The module connects via the MIPI CSI-2 interface, and image capture is managed through software interfaces provided by the Raspberry Pi operating system.

To simulate real-world infrastructure deployments, Roadside Units (RSUs) manufactured by Movyon Electronics have been used. These units are industrial-grade devices that fully comply with the European ITS-G5 (IEEE 802.11p) standards and the C-V2X specifications defined in 3GPP Releases 14 and 15. Each RSU integrates a dedicated ITS-G5 software stack and operates within the regulated 5.9 GHz ITS band, providing reliable V2X communications with a reception sensitivity of -97 dBm. Accurate geolocation is ensured by multi-constellation GNSS support (GPS, Galileo, GLONASS, and BeiDou), enabling precise positioning for cooperative awareness and infrastructure-assisted applications. The devices are CE RED certified and housed in IP67 enclosures, ensuring suitability for outdoor operation.

During video experiments, the on-board unit captures video at 1280×720 pixels at 30 frames per second and the stream is encapsulated in MPEG-TS and sent over UDP, while individual frames are encoded in M-JPEG. This solution was adopted to keep computational requirements within the platform's limits; alternative approaches based on software encoding were found to impose substantially higher CPU load. Although M-JPEG is less bandwidth-efficient than inter-frame codecs, it enables stable operation while preserving a level of visual quality suitable for remote-driving scenarios. The transmission chain was also tuned to minimize buffering, thereby keeping end-to-end latency low.

\section{Software Setup}
\label{sec_testbed}
\begin{figure}[h!]
    \centering
    \includegraphics[width=\textwidth]{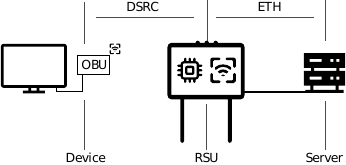}
    \caption{OBU to RSU/OBU setup.}
    \label{fig:setup}
\end{figure}

This work contains two different communication paradigms: OBU-to-OBU and OBU-to-RSU (Figure \ref{fig:setup}). In the OBU-to-OBU configuration, a secondary OBU unit was repurposed to emulate the Road-Side Unit (RSU) functionality, facilitating the analysis of vehicle-to-infrastructure and inter-vehicle communication. This scenario was evaluated using a comprehensive protocol suite across both IPv4 unicast and IPv6 multicast address families to quantify the performance implications of different network-layer implementations. In contrast, in the OBU-to-RSU configuration with a dedicated commercial RSU, communication was limited only to IPv6 multicast protocols due to device limitations.

Regarding the video transmission subsystem, the acquisition pipeline was orchestrated using a Raspberry Pi Camera module to capture raw sensor data. The subsequent encoding and transmission stages were managed using the \FFmpeg application, which was tuned to minimize end-to-end latency. This configuration used the MPEG-TS container format over UDP, using a custom parameter set that prioritizes frame accuracy and temporal efficiency. To rigorously evaluate the impact of network congestion on video quality and transmission stability, multiple different bitrate configurations were systematically tested. For baseline network throughput and bandwidth characterization, independent tests were conducted using raw UDP traffic generated with the \iPerf tool; in this context, a spectrum of varying bandwidth constraints was applied to characterize the link's capacity across different saturation levels. 

Complementing these software-level assessments, a series of physical layer modifications was implemented in all test scenarios to evaluate the impact of wireless link characteristics on transmission reliability. Specifically, two different queue discipline algorithms were investigated: the CAKE
(Common Applications Kept Enhanced) queue discipline \cite{8475045}, which combines fair queuing with active queue management mechanisms derived from FQ-CoDel, and the standard default queueing mechanism without such optimization. Furthermore, three different Modulation and Coding Schemes (MCS) were systematically varied to assess the trade-offs between data rate and link robustness: Binary Phase Shift Keying (BPSK), Quadrature Phase Shift Keying (QPSK), and 16-Quadrature Amplitude Modulation (16-QAM) (Table \ref{tab:modulation_ids}). 

To set up the designated MCS, regarding IPv4 cases, the following command has been launched:
\lstset{basicstyle=\ttfamily, breaklines=true, columns=flexible}
\begin{lstlisting}
echo <ID> > /sys/kernel/debug/ieee80211/phy0/rc/fixed_rate_idx
\end{lstlisting}
where {$\tt ID$} is the selected MCS index. Concerning, on the other hand, IPv6 scenarios, the MCS can be changed using the command:
\begin{lstlisting}
iw dev wlan0 set mcast_rate <rate>
\end{lstlisting}
where {$\tt rate$} is the maximum rate allowed, which implicitly specifies the MCS used (only 6, 9, or 12 are allowed).

\begin{figure*}[t]
    \centering
    \begin{subfigure}{0.32\textwidth}
        \centering
         \includegraphics[width=\textwidth]{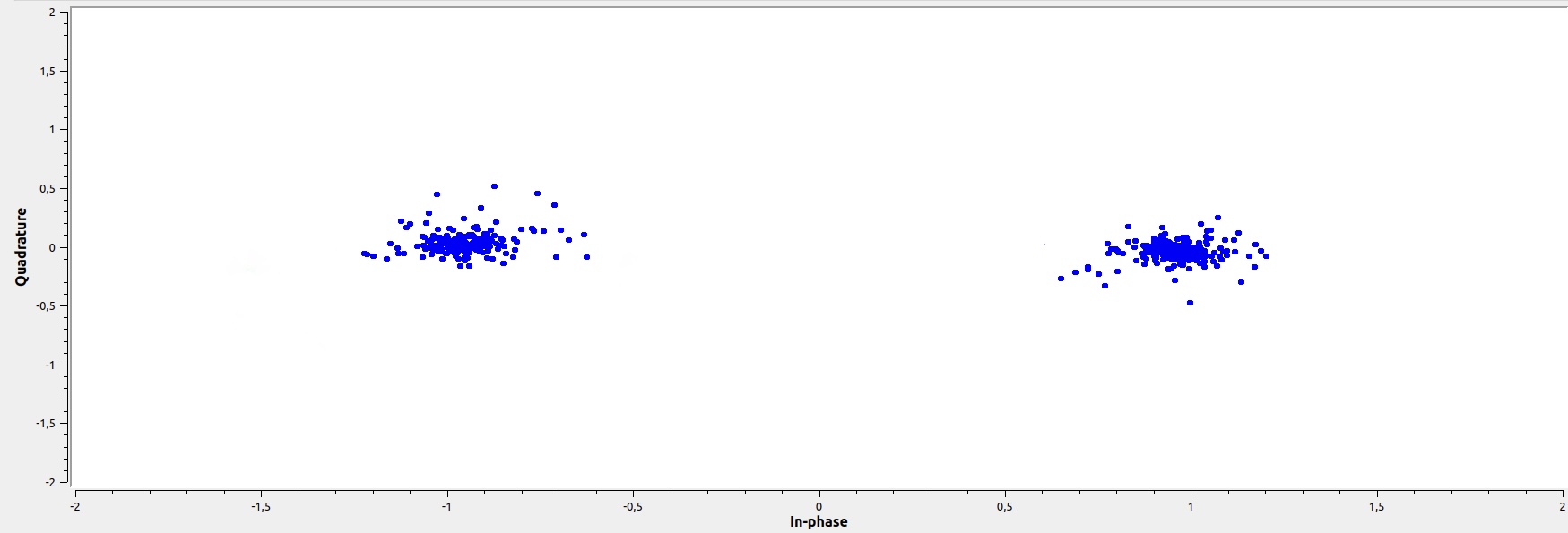} 
        \caption{BPSK (\textit{IPv4 \& IPv6})}
        \label{fig:sdr_bpsk}
    \end{subfigure}
    \hfill
    \begin{subfigure}{0.32\textwidth}
        \centering
         \includegraphics[width=\textwidth]{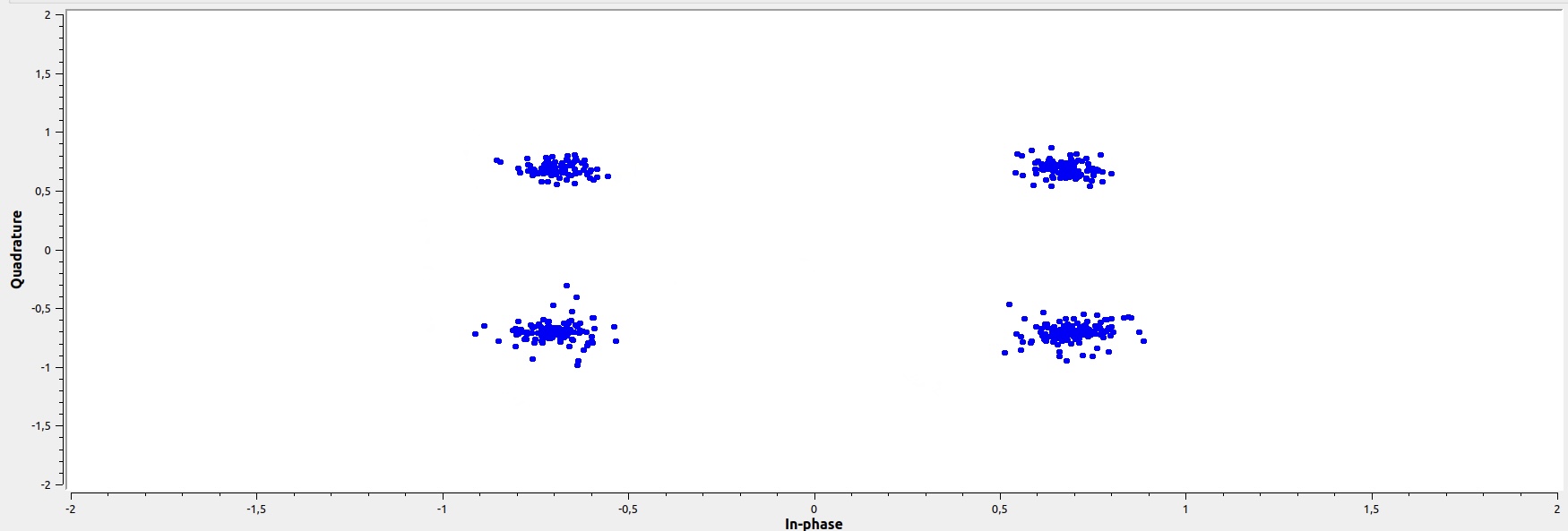} 
        \caption{QPSK (\textit{IPv4 \& IPv6})}
        \label{fig:sdr_qpsk}
    \end{subfigure}
    \hfill
    \begin{subfigure}{0.32\textwidth}
        \centering
        \includegraphics[width=\textwidth]{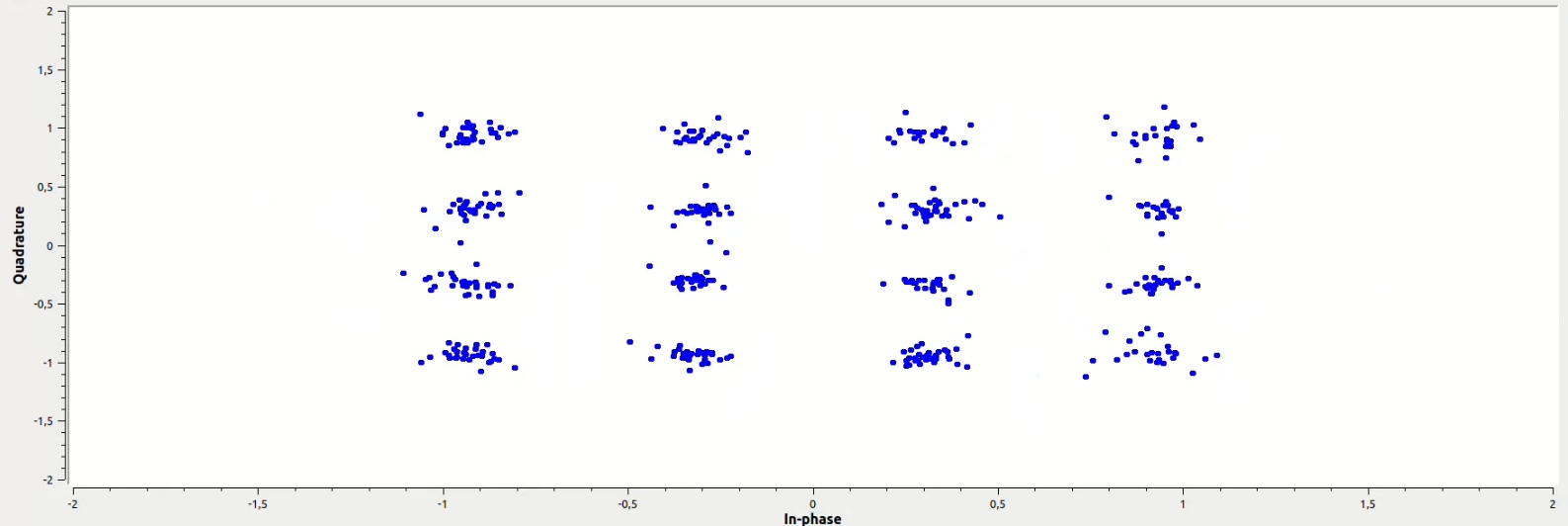}
        \caption{16QAM (\textit{IPv4 only})}
        \label{fig:sdr_16qam}
    \end{subfigure}
    \caption{SDR}
    \label{fig:sdr}
\end{figure*}

Crucially, SDR equipment was used to observe and validate the transmitted modulation characteristics, ensuring that the configured MCS values were correctly applied by the wireless interface, in order to ensure that the theoretical modulation parameters were accurately performed in the physical transmission chain (Figures \ref{fig:sdr_bpsk}, \ref{fig:sdr_qpsk} and \ref{fig:sdr_16qam}).

\begin{table}[htbp]
    \centering
    \small 
    \setlength{\tabcolsep}{12pt} 
    \begin{tabular}{lcS[table-format=2.1]} 
    \toprule
    \textbf{ID} & \textbf{MCS} & {\textbf{Bandwidth (Mbps)}} \\
    \midrule
    0  & BPSK     & 2.4 \\
    1  & QPSK 1/2 & 4.4 \\
    2  & QPSK 3/4 & 6.0 \\
    3  & 16QAM    & 7.3 \\
    4  & 16QAM    & 9.4 \\
    5  & 16QAM    & 11.0 \\
    6  & 16QAM    & 11.5 \\
    7  & 16QAM    & 12.0 \\
    \midrule
    8  & BPSK     & 4.3 \\
    9  & QPSK 1/2 & 7.1 \\
    10 & QPSK 3/4 & 9.3 \\
    11 & 16QAM    & 10.7 \\
    12 & 16QAM    & 13.0 \\
    13 & 16QAM    & 14.0 \\
    14 & 16QAM    & 13.9 \\
    15 & 16QAM    & 8.0 \\
    \midrule
    \dots & \dots & {\dots} \\
    23 & 16QAM    & 10.0 \\
    \bottomrule
    \end{tabular}
    \caption{Modulation IDs and corresponding bandwidths.}
    \label{tab:modulation_ids}
\end{table}

\section{Tests and Results}
\label{sec_test_results}
The following subsections evaluate latency characteristics in IEEE
802.11p networks, considering both physical-layer parameters and
network-layer configurations. Initially, the analysis examines UDP latency across varying traffic loads, establishing a baseline comparison between IPv4 unicast and IPv6 multicast under different modulation and coding schemes (MCS). This assessment is then extended to a video streaming scenario to evaluate real-world application performance, specifically investigating the impact of video bitrates and the effectiveness of the CAKE (Common Applications Kept Enhanced) active queue management (AQM) algorithm. Each test was performed multiple times, in an indoor test area, lasting 60 seconds each; no differences were observed across runs. As described above, the OBU-to-OBU and OBU-to-RSU paradigms were used (for the IPv6 scenario), but no performance differences were observed.

\subsection{UDP Traffic}
\begin{figure}
    \subfloat[$2 Mbps$ UDP]{
        \includegraphics[width=0.8\linewidth]{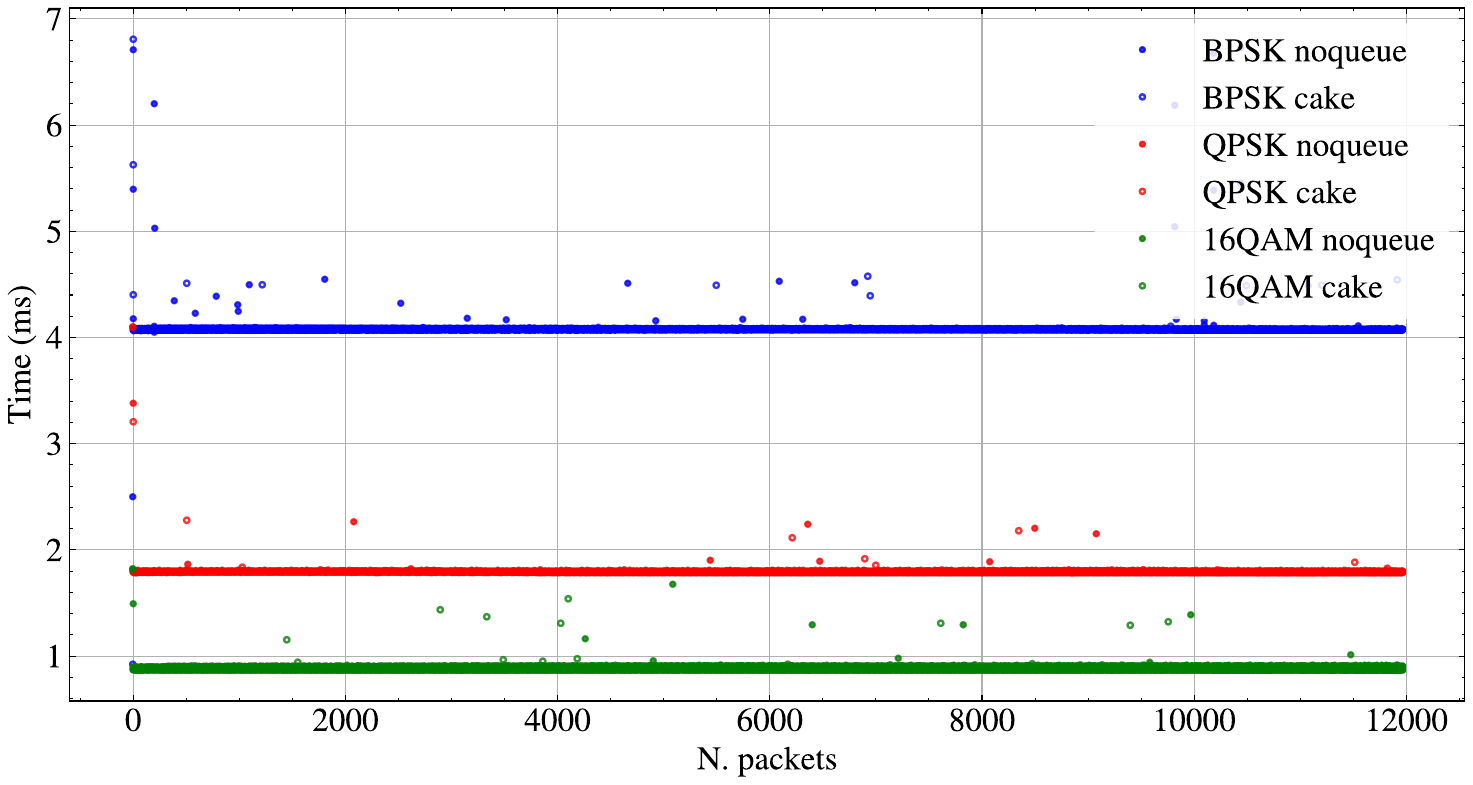}
    \label{fig:udp2_ipv4_ucast}
    }
    \hfill
    \subfloat[$4 Mbps$ UDP]{
    \includegraphics[width=0.8\linewidth]{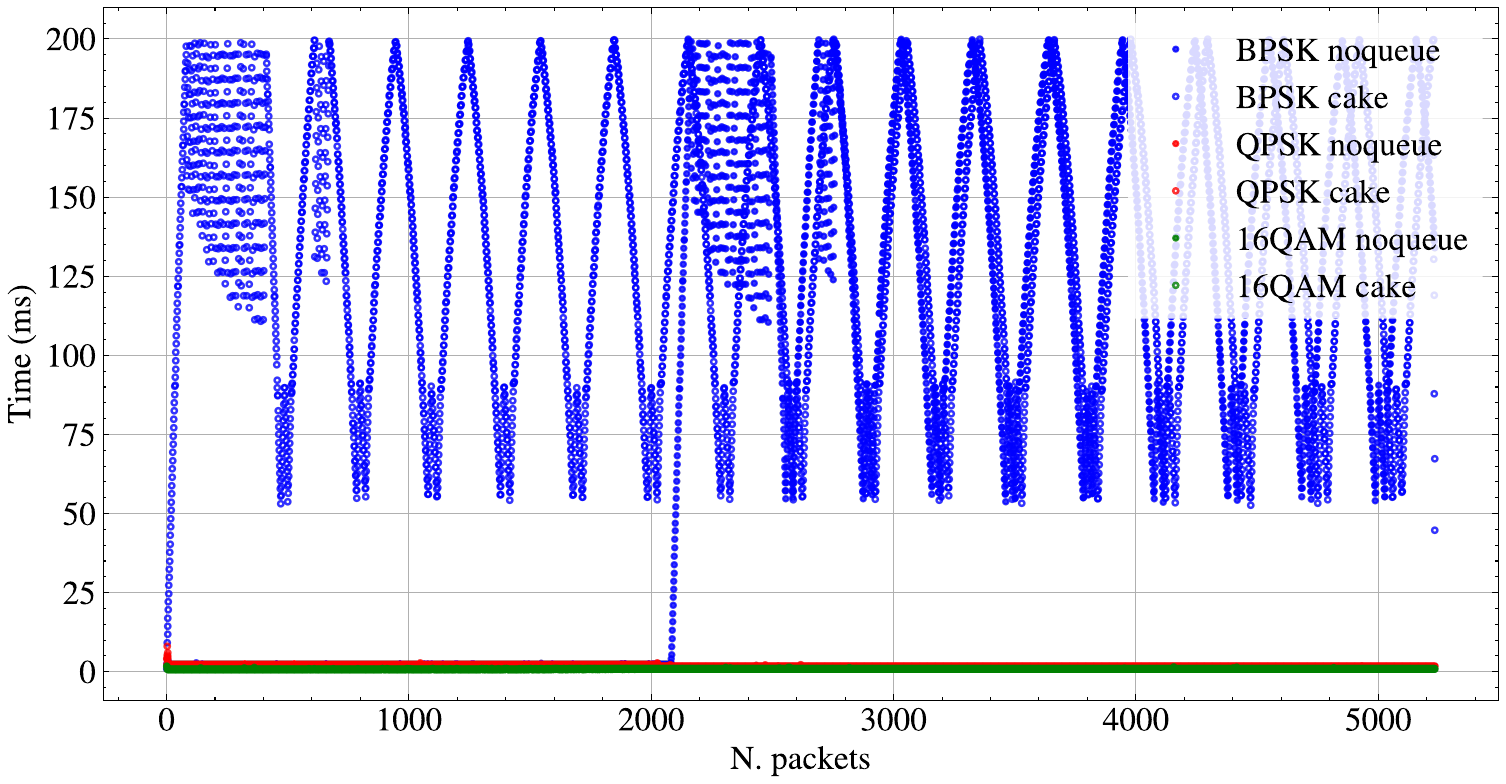}
    \label{fig:udp4_ipv4_ucast}
    }
    \hfill
    \subfloat[$20 Mbps$ UDP]{
    \includegraphics[width=0.8\linewidth]{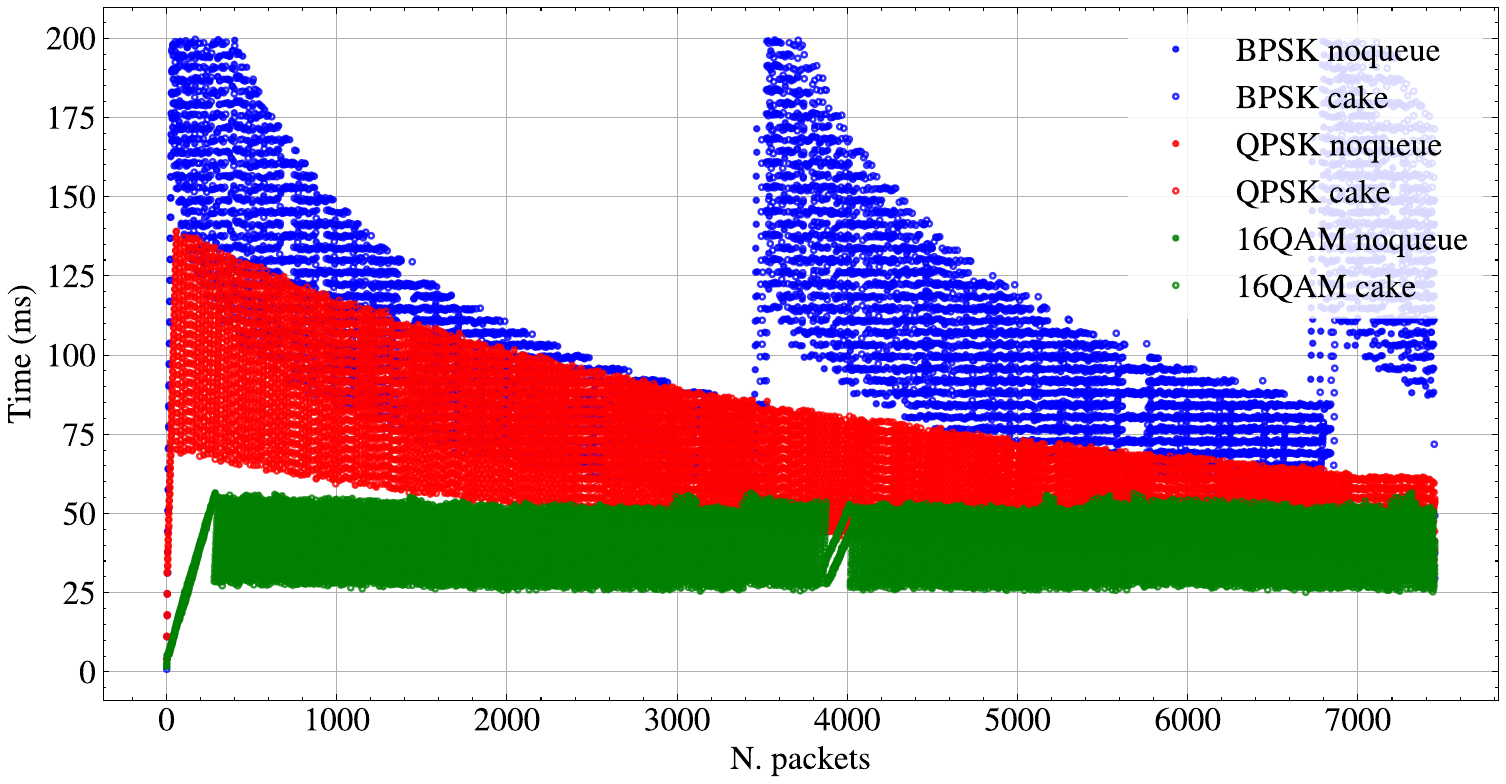}
    \label{fig:udp20_ipv4_ucast}
    }
    \caption{IPv4 unicast comparison UDP.}
    \label{fig:udp_ipv4_ucast}
\end{figure}

\begin{figure}
    \subfloat[$2 Mbps$ UDP]{
        \includegraphics[width=0.8\linewidth]{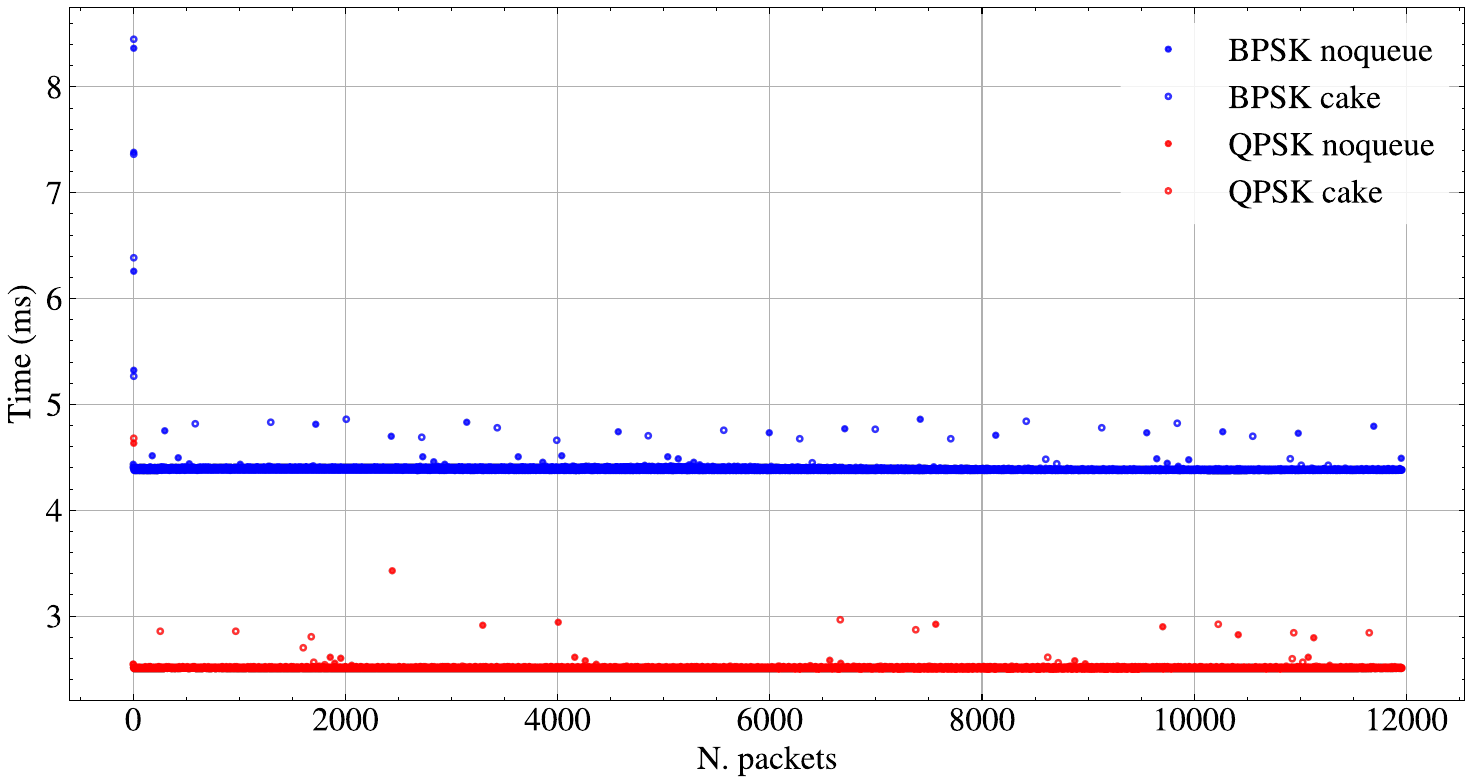}
    \label{fig:udp2_ipv6_multi}
    }
    \hfill
    \subfloat[$4 Mbps$ UDP]{
    \includegraphics[width=0.8\linewidth]{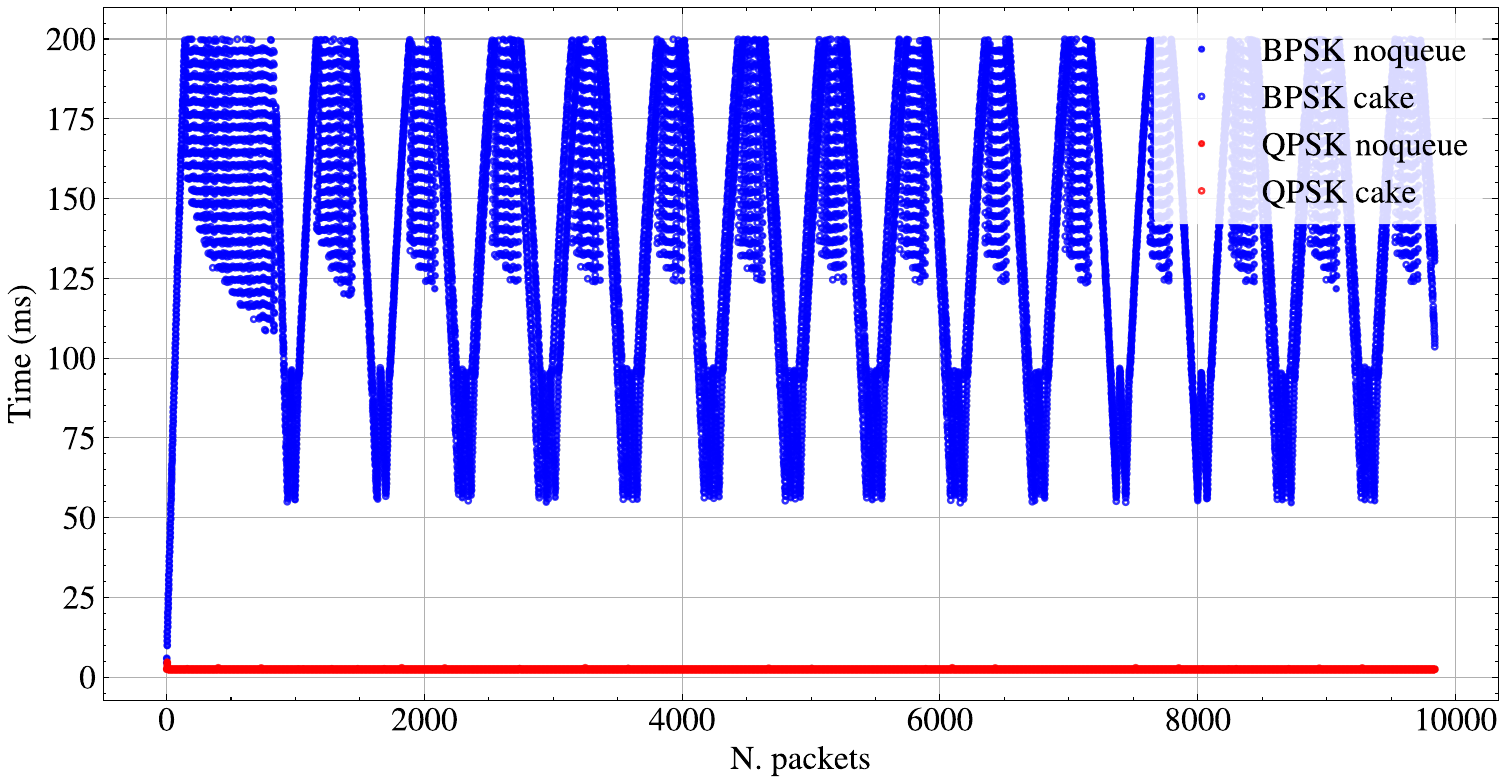}
    \label{fig:udp4_ipv6_multi}
    }
    \hfill
    \subfloat[$20 Mbps$ UDP]{
    \includegraphics[width=0.8\linewidth]{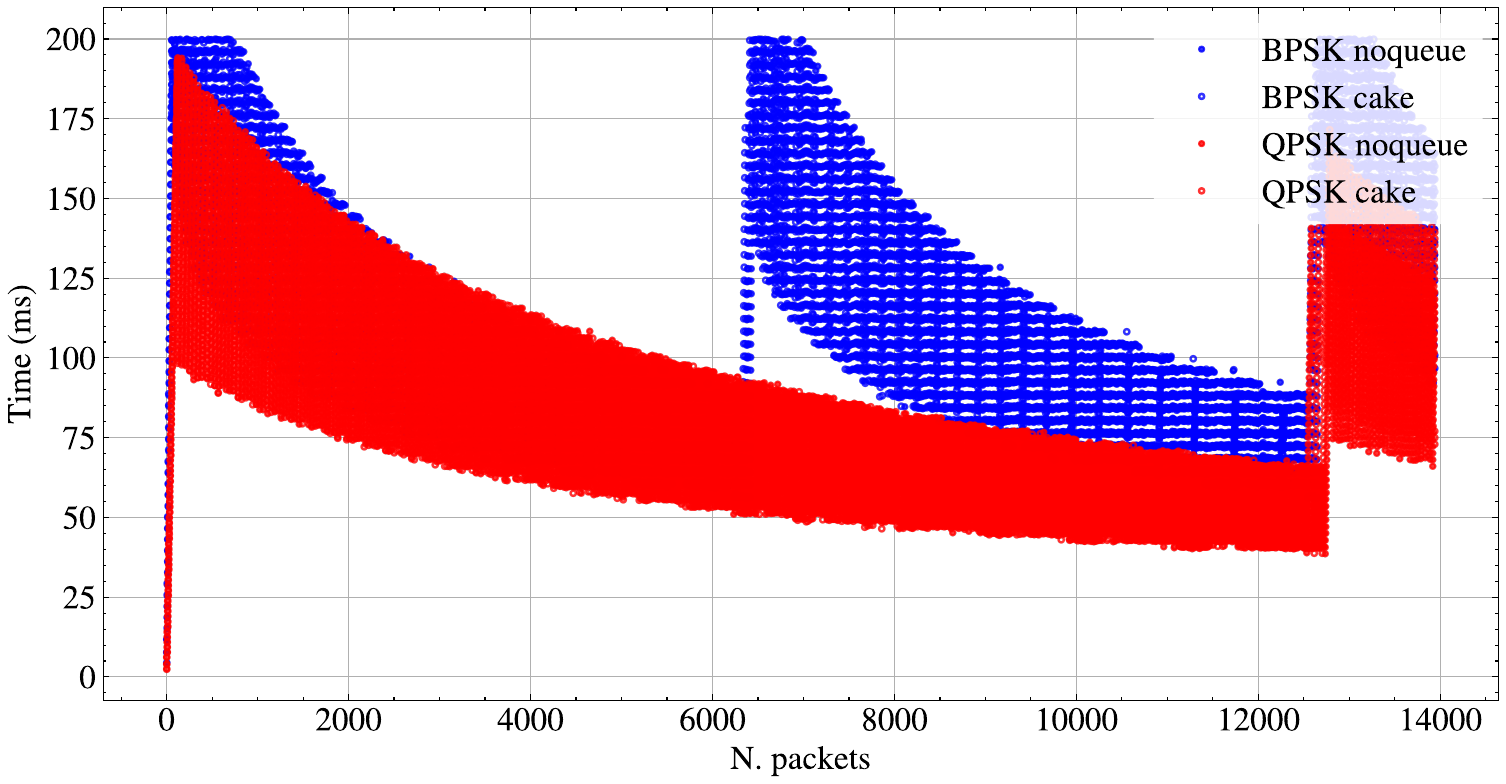}
    \label{fig:udp20_ipv6_multi}
    }
    \caption{IPv6 multicast comparison UDP.}
    \label{fig:udp_ipv6_multi}
\end{figure}

Figures \ref{fig:udp_ipv4_ucast} and \ref{fig:udp_ipv6_multi} provide a comprehensive comparison of UDP latency over IEEE 802.11p under different traffic loads (2, 4, and 20 Mbps), highlighting the combined impact of IP version, transmission mode, and modulation scheme. The results indicate that the CAKE does not introduce a measurable latency penalty; the observed results for both configurations appear closely aligned, with no significant differences.

At low load (2 Mbps), both IPv4 unicast (Fig. \ref{fig:udp2_ipv4_ucast}) and IPv6 multicast (Fig. \ref{fig:udp2_ipv6_multi}) show stable and nearly constant delay. In this regime, latency is mainly determined by the modulation scheme. BPSK exhibits the highest delay, QPSK shows intermediate values, and higher-order modulations such as 16QAM (available only in the IPv4 unicast scenario) achieve the lowest latency. The differences between IPv4 and IPv6 are limited, although IPv6 multicast exhibits slightly higher delay variability.

At 4 Mbps (Figs. \ref{fig:udp4_ipv4_ucast} and \ref{fig:udp4_ipv6_multi}), the system enters a transitional regime and delay oscillations become visible. This behavior is particularly evident for BPSK, whose low physical data rate leads to queue buildup. The effect is more pronounced in IPv6 multicast (Fig. \ref{fig:udp4_ipv6_multi}), where the lack of link-layer retransmission and reliability mechanisms in multicast transmissions increases delay variability. In contrast, IPv4 unicast (Fig. \ref{fig:udp4_ipv4_ucast}) shows better stability for QPSK and 16QAM.

Under high load (20 Mbps), the network becomes clearly congestion-dominated and latency is mainly driven by queueing and channel contention. In IPv4 unicast (Fig. \ref{fig:udp20_ipv4_ucast}), BPSK and QPSK initially experience large delay values, followed by a gradual decrease as the queues stabilize. In contrast, 16QAM maintains significantly lower and more stable latency thanks to its higher data rate. The abrupt truncation of the delay curves for BPSK likely indicates increased packet loss, as packets may be dropped before being measured due to buffer exhaustion.

IPv6 multicast (Fig. \ref{fig:udp20_ipv6_multi}) shows higher variability and less predictable behavior under the same load. Delay values are more dispersed and regime shifts become visible, particularly for BPSK. This behavior is consistent with the lack of retransmission mechanisms and the higher sensitivity of multicast traffic to congestion.

Overall, the results indicate that modulation order dominates latency at low traffic loads. As congestion increases, the choice of transmission mode becomes more relevant. IPv4 unicast exhibits more stable and predictable latency than IPv6 multicast, whereas higher-order modulations significantly improve delay performance under heavy-traffic conditions.

\subsection{Video Streaming}
\begin{figure}
    \subfloat[$2 Mbps$ stream video]{
    \includegraphics[width=0.45\linewidth]{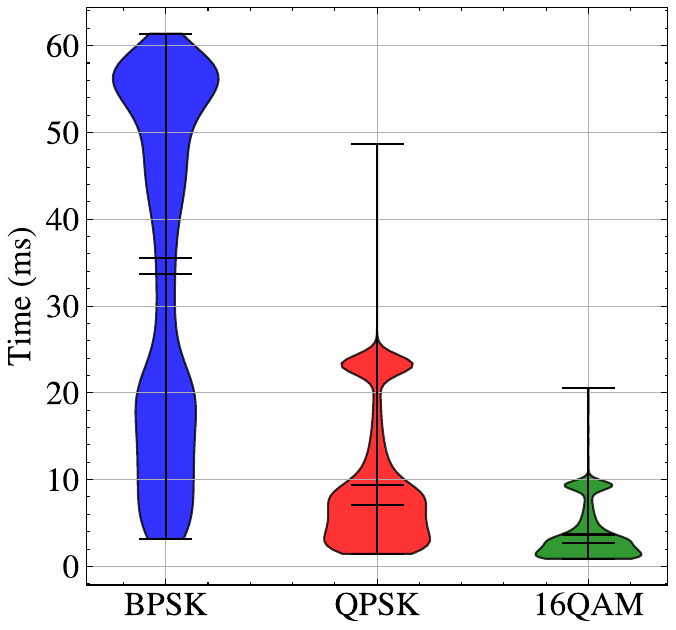}
    \label{fig:noqueue_ipv4_ucast(a)}
    }
    \hfill
    \subfloat[$4 Mbps$ stream video]{
    \includegraphics[width=0.45\linewidth]{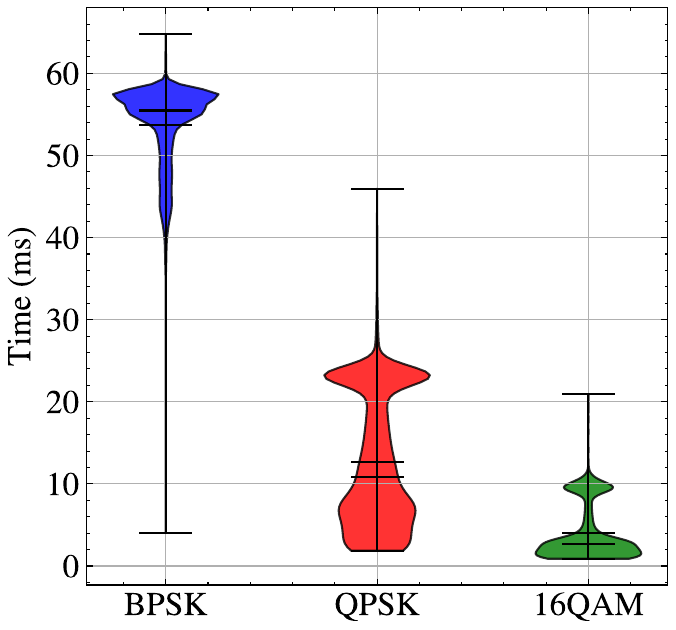}
    \label{fig:noqueue_ipv4_ucast(b)}
    }
    \caption{Stream video IPv4 unicast comparison noqueue.}
    \label{fig:noqueue_ipv4_ucast}
\end{figure}

\begin{figure}
    \subfloat[$2 Mbps$ stream video]{
    \includegraphics[width=0.45\linewidth]{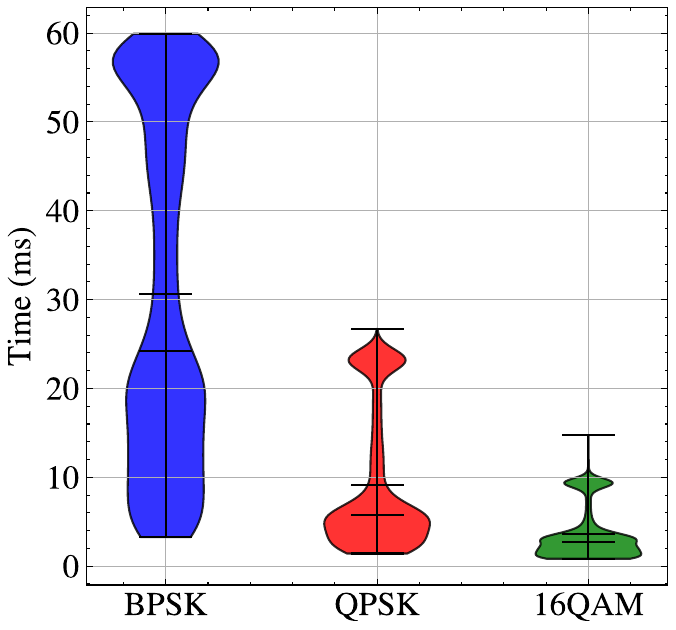}
    \label{fig:cake_ipv4_ucast(a)}
    }
    \hfill
    \subfloat[$4 Mbps$ stream video]{
    \includegraphics[width=0.45\linewidth]{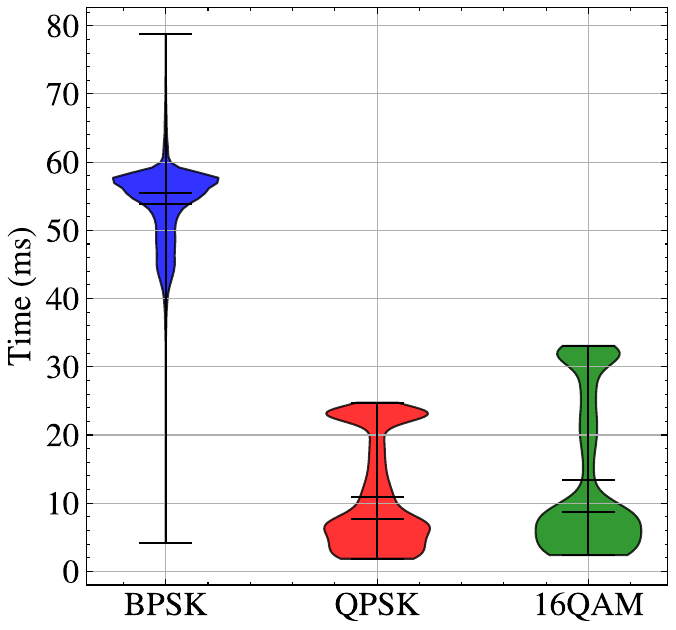}
    \label{fig:cake_ipv4_ucast(b)}
    }
    \caption{Stream video IPv4 unicast comparison cake.}
    \label{fig:cake_ipv4_ucast}
\end{figure}

\begin{figure}
    \subfloat[$2 Mbps$ stream video]{
    \includegraphics[width=0.45\linewidth]{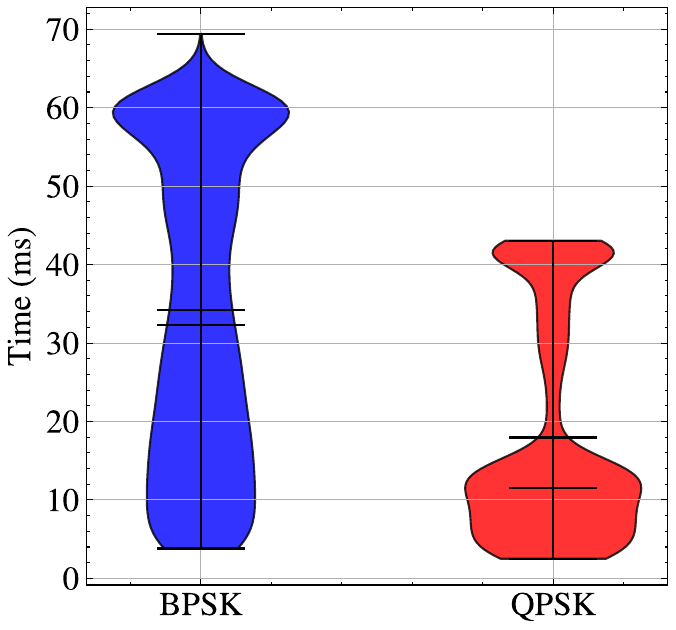}
    \label{fig:noqueue_ipv6_multi(a)}
    }
    \hfill
    \subfloat[$4 Mbps$ stream video]{
    \includegraphics[width=0.45\linewidth]{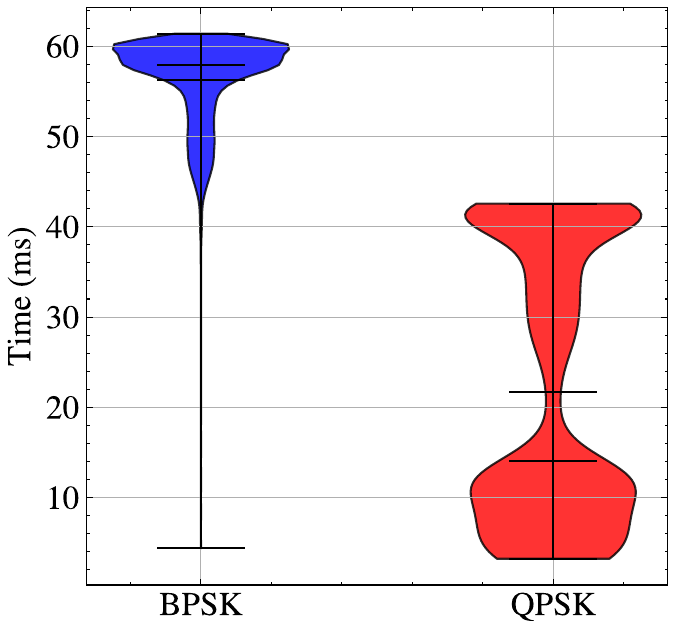}
    \label{fig:noqueue_ipv6_multi(b)}
    }
    \caption{Stream video IPv6 multicast comparison noqueue.}
    \label{fig:noqueue_ipv6_multi}
\end{figure}

\begin{figure}
    \subfloat[$2 Mbps$ stream video]{
    \includegraphics[width=0.45\linewidth]{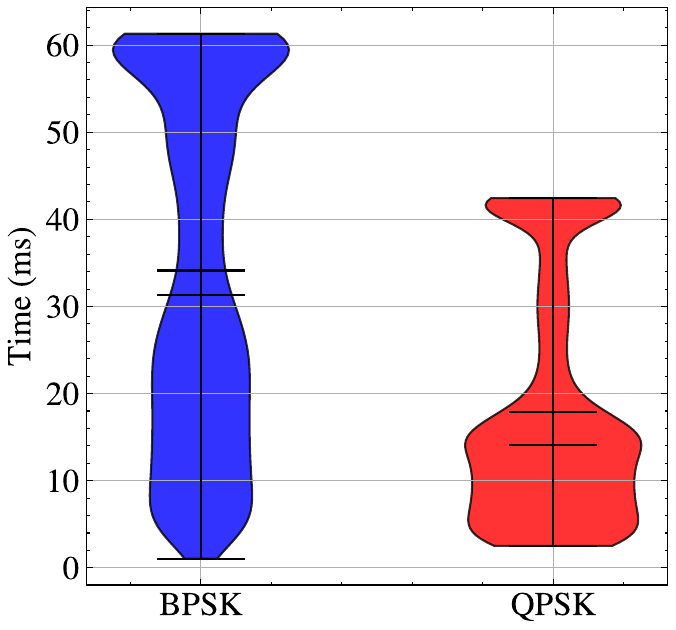}
    \label{fig:cake_ipv6_multi(a)}
    }
    \hfill
    \subfloat[$4 Mbps$ stream video]{
    \includegraphics[width=0.45\linewidth]{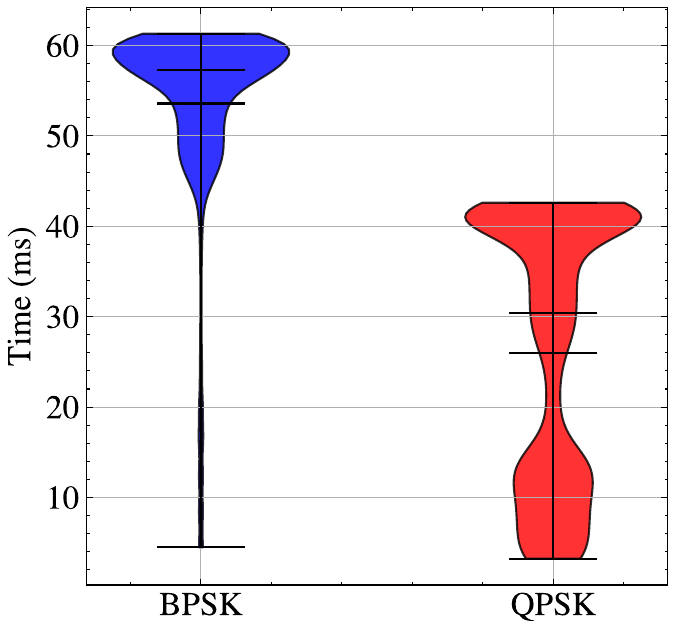}
    \label{fig:cake_ipv6_multi(b)}
    }
    \caption{Stream video IPv6 multicast comparison cake.}
    \label{fig:cake_ipv6_multi}
\end{figure}

The results reported in Figs. \ref{fig:noqueue_ipv4_ucast}, \ref{fig:cake_ipv4_ucast}, \ref{fig:noqueue_ipv6_multi}, and \ref{fig:cake_ipv6_multi} illustrate the combined impact of modulation scheme, traffic configuration, network protocol, and video bitrate on delay performance in IEEE 802.11p video streaming scenarios.

In the IPv4 unicast configuration without queue management (Fig. \ref{fig:noqueue_ipv4_ucast}), the modulation scheme strongly affects delay behavior. BPSK exhibits the highest latency and variability, with wide delay distributions and long tails exceeding 50–60 ms. QPSK significantly reduces both median delay and dispersion. The best performance is obtained with 16QAM, which maintains the lowest and most stable delay values, typically within a few milliseconds. Increasing the video bitrate from 2 Mbps to 4 Mbps slightly increases delay variability but does not change the relative performance among modulation schemes.

Introducing CAKE in the IPv4 scenario (Fig. \ref{fig:cake_ipv4_ucast}) does not significantly alter delay behavior. The delay distributions obtained with and without CAKE are very similar, suggesting that active queue management has limited impact under these conditions.

A similar trend can be observed in the IPv6 multicast configuration without queue management (Fig. \ref{fig:noqueue_ipv6_multi}). BPSK again exhibits the highest variability and peak delays. QPSK exhibits greater dispersion than in the IPv4 unicast case, especially at 4 Mbps. This behavior reflects the reduced MAC-layer efficiency of multicast transmission and its higher sensitivity to congestion.

When CAKE is enabled in the IPv6 configuration (Fig. \ref{fig:cake_ipv6_multi}), no clear reduction in delay variability is observed; the delay distributions remain almost unchanged for both tested modulation schemes. This suggests that queue management alone cannot compensate for the limitations introduced by low-rate modulation and multicast delivery.

Overall, the results confirm that the modulation scheme is the primary factor affecting delay performance in IEEE 802.11p video streaming. BPSK consistently exhibits the worst-case delay, whereas higher-order modulations yield significantly lower and more stable delays. Higher video bitrates slightly increase delay dispersion. In addition, IPv6 multicast is more sensitive to congestion than IPv4 unicast, while CAKE does not provide measurable improvements in delay stability under the tested conditions.

\section{Conclusions}
\label{sec_conclusions}
This paper presents an experimental evaluation of UDP transmission over IEEE 802.11p using a real ITS-G5 testbed comprising Raspberry Pi–based OBUs and commercial RSUs. The experiments analyzed the impact of different modulation schemes, transmission modes, and network-layer configurations on latency performance.

The results show that the modulation scheme is the main factor affecting delay under low traffic load. Higher-order modulations significantly reduce latency and delay variability compared with BPSK. As traffic load increases, the choice of transmission mode becomes more relevant. In particular, IPv4 unicast provides more stable and predictable delay than IPv6 multicast, which shows higher variability under congestion. The evaluation of CAKE queue management did not reveal measurable improvements in delay stability in the tested scenarios. This suggests that queue management is limited in effectiveness in this context, as the primary bottleneck lies in the physical layer and transmission characteristics rather than in buffer-induced queuing.

In general, these findings indicate that, in the ITS-G5 scenarios considered, optimizing PHY-layer parameters and transmission strategies appears more effective than relying solely on active queue management mechanisms. Future work will investigate more complex and dense vehicular scenarios, including heterogeneous traffic patterns and adaptive transmission strategies, where queue management techniques such as CAKE may play a more significant role.
%Future work will investigate Quality of Service mechanisms based on EDCA access categories and their interaction with queue management in heterogeneous traffic scenarios. In addition, adaptive mechanisms for dynamic MCS selection based on channel conditions will be explored to improve performance in dense vehicular environments.

\balance
\bibliographystyle{IEEEtran}
\bibliography{bibliography}

\end{document}